\documentclass{article}
\usepackage{ijcai22}

\usepackage{times}

\usepackage{soul}
\usepackage{url}
\usepackage[hidelinks]{hyperref}
\usepackage[utf8]{inputenc}
\usepackage[small]{caption}
\usepackage{graphicx}
\usepackage{amsmath}
\usepackage{amsfonts}
\usepackage{booktabs}
\usepackage{cleveref}
\usepackage{latexsym}
\usepackage{multirow}
\usepackage{hyperref}
\hypersetup{colorlinks,allcolors=black}

\usepackage{xcolor}

\newcommand{\citet}[1]{\citeauthor{#1}~\shortcite{#1}}

\title{Model-based Multi-agent Reinforcement Learning: \\ Recent Progress and Prospects}

\author{
Xihuai Wang\and
Zhicheng Zhang\and
Weinan Zhang
\affiliations
Shanghai Jiao Tong University\\
\emails
\{leoxhwang, zhangzhicheng1, wnzhang\}@sjtu.edu.cn
}

\begin{document}

\maketitle

\begin{abstract}
Significant advances have recently been achieved in Multi-Agent Reinforcement Learning (MARL) which tackles sequential decision-making problems involving multiple participants. However, MARL requires a tremendous number of samples for effective training. On the other hand, model-based methods have been shown to achieve provable advantages of sample efficiency. However, the attempts of model-based methods to MARL have just started very recently. This paper presents a review of the existing research on model-based MARL, including theoretical analyses, algorithms, and applications, and analyzes the advantages and potential of model-based MARL. Specifically, we provide a detailed taxonomy of the algorithms and point out the pros and cons for each algorithm according to the challenges inherent to multi-agent scenarios. We also outline promising directions for future development of this field.

\end{abstract}

\section{Introduction}


Recent years have witnessed the tremendous success of Multi-Agent Reinforcement Learning (MARL) in various complex tasks involving multiple participants, including playing real-time strategy games~\cite{DBLP:conf/gecco/ArulkumaranCT19,DBLP:conf/aaai/YeLSSZWYYWGCYZS20}, card games~\cite{DBLP:journals/ai/BardFCBLSPDMHDM20}, sports games~\cite{DBLP:conf/aaai/KurachRSZBERVMB20}, autonomous driving~\cite{DBLP:journals/corr/abs-2010-09776}, and multi-robot navigation~\cite{DBLP:conf/icra/LongFLLZP18}. MARL solves the sequential decision-making problem in which multiple agents interact in a common environment, receive reward signals, and improve their policies to maximize cumulative reward. 

Despite the success stories of MARL algorithms, several challenges of multi-agent scenarios give rise to the sample inefficiency of MARL. Specifically, the major challenge is the non-stationarity, which means that all agents improve their policies according to their observations and rewards, leading to a non-stationary environment from the perspective of an individual agent. The guarantees of the effectiveness of most reinforcement learning algorithms may not hold in multi-agent scenarios since the Markov property to which the guarantees are tied could be invalidated in non-stationary multi-agent scenarios. Furthermore, the dimensions of the variable spaces could grow exponentially w.r.t. the number of agents, leading to the dimension blowup in the policy search space~\cite{DBLP:conf/atal/Hernandez-LealK20}. Due to two significant challenges, together with the associated problems such as partial observability, coordination, credit assignment, and scalability, MARL requires a tremendous number of samples for effective training~\cite{gronauer2021multi}.

On the other hand, model-based methods in single-agent RL scenarios have shown their advantages in sample efficiency both practically~\cite{DBLP:journals/corr/abs-1907-02057} and theoretically~\cite{DBLP:conf/colt/SunJKA019}. However, model-based MARL, in the sense that agents learn environment models\footnote{In this paper, an environment model refers to a dynamics model and a reward function model.} and leverage the learned model for policy improvement, has just started to attract attention and to be applied in real-world tasks~\cite{DBLP:conf/aiide/ChockalingamSBG18}. Moreover, simply applying off-the-shelf model-based RL methods in single-agent scenarios to multi-agent scenarios would hardly work due to the previously mentioned challenges.

This paper provides an overview of existing model-based MARL methods and shows the potential of model-based MARL methods in improving sample efficiency and resolving the inherent challenges. First, we present the theoretical analyses on the sample efficiency of model-based MARL in limited settings. Second, we give a taxonomy of the algorithms, organized by the training schemes, the opponent awareness, and the environment model usage. Each algorithm is then discussed in detail on its pros and cons according to the environment usage and whether the model-based methods resolve or sharpen the challenges. Finally, we spotlight several areas worth exploring and outline future directions that have the potential to the development of this field based on our  knowledge.





\section{Background}
\subsection{Problem Formulation}
The sequential decision making problem in multi-agent scenarios is generally formulated as a stochastic game~\cite{shapley1953stochastic}, which is also known as Markov game. An $n$-agent stochastic game can be formalized as a tuple $(\mathcal{S}, \{\mathcal{A}^{i}\}_{i \in \mathcal{N}}, \{R^{i}\}_{i \in \mathcal{N}}, \mathcal{T}, \gamma)$, where $\mathcal{N} = \{1, \ldots, n\}$ is the set of agents, $\mathcal{S}$ is the state space of the stochastic game, $\mathcal{A}^{i}$ is the action space of agent $i$. Denote the joint action space as $\mathcal{A}=\mathcal{A}^{1}\times \cdots \times \mathcal{A}^{n}$, $R^{i}: \mathcal{S}\times \mathcal{A} \mapsto  \mathbb{R}$ is the reward function of agent $i$, $\mathcal{T}: \mathcal{S}\times \mathcal{A} \mapsto \Delta(\mathcal{S})$ is the dynamics function denoting the transition probability to each state. $\gamma \in [0, 1)$ is the discount factor for future rewards.

At each timestep $t$, each agent $i$ takes action $a^{i}_{t}$ simultaneously according to the state $s_t$. Then the state transits to $s_{t+1}$ and each agent $i$ receives reward $r^{i}_{t}$. Denoting the joint action of other agents as $a^{-i}_{t}=\{a_{t}^{j}\}_{j\neq i}$, we can formulate the joint policy of other agents from the perspective of agent $i$ as $\pi^{-i}(a^{-i}_{t}|s_{t})=\Pi_{j\in\{-i\}}\pi^{j}(a^{j}_{t}|s_{t})$, where $\pi^{j}: \mathcal{S}\mapsto \Delta(\mathcal{A}^{i})$ is the policy of agent $j$. Each agent $i$ aims at finding its optimal policy to maximize the expected return (cumulative reward), defined as
\begin{align*}
    \pi_i^* &= \operatorname{argmax}_{\pi^i} \eta_i[\pi^{i}, \pi^{-i}] \\
    &=\operatorname{argmax}_{\pi^i} \mathbb{E}_{\tau \sim (\mathcal{T},\pi^{i},\pi^{-i})}
    \Big[\sum_{t=0}^\infty \gamma^{t}R^{i}(s_{t}, a_{t}^{i}, a_{t}^{-i}) \Big]~,
\end{align*}
where $\tau = \{(s_{0}, a_{0}^{i}, a_{0}^{-i}), (s_{1}, a_{1}^{i}, a_{1}^{-i}), \ldots \}$ denotes the sampled trajectory.

From the objective, we note that the optimal performance and the optimal policy of agent $i$ depend not only on its own policy but also on the behaviors of other agents.

An important motivation for model-based methods is to reduce the sample complexity, i.e., how many samples are needed to achieve a certain performance in a multi-agent environment. Following~\citet{DBLP:conf/ijcai/0001WSZ21}, from the perspective of an arbitrary agent in this stochastic game, we define the sample complexity in terms of two parts: (i) \textit{dynamics sample complexity}, i.e., the number of state samples from the dynamics environment in which the group of agents interact, and (ii) \textit{opponent sample complexity}, i.e., the total number of opponent action samples, including the samples from real interactions with the environment and the number of actions requested through communication in decentralized settings.

\subsection{Challenges in MARL}
\label{subsec: challenge}
To reveal the role of model-based methods in MARL, we depict several challenges~\cite{DBLP:journals/air/DuD21,DBLP:journals/corr/abs-1911-10635,DBLP:journals/corr/abs-1812-11794} that MARL suffers from and analyze whether current model-based methods alleviate or sharpen these challenges in \Cref{sec:alg}.

\paragraph{Non-stationarity.} One of the key challenges of MARL origins from the fact that multiple agents interact with the environment to improve their policies simultaneously and individually. From the perspective of a single agent, the environment becomes non-stationary due to the co-adaption of the opponents, i.e., the perceived transition function and reward function change over time. The policy learned in the non-stationary environment has mismatched expectation about the opponents' policies, and the optimal behavior of the agent depends not only on the perceived information but also on the behavior of its opponents. Thus, the Markov assumption is violated, and the theoretical guarantees for single-agent scenarios may be lost in multi-agent systems. We refer the readers to~\citet{DBLP:journals/corr/abs-1906-04737} for further discussions on the non-stationarity.

\paragraph{Partial Observability.} Unlike the stochastic game setting, agents may neither observe the global state of the environment nor have full knowledge about opponents in real-world applications. The Markov property may not be satisfied under partially observable scenarios. Agents need to approximate the global environment state based on limited knowledge to alleviate the non-Markov problem. The decision-making problem in partially observable scenarios has a similar definition as a stochastic game, but with observation spaces $\{\Omega^{i}\}_{i \in \mathcal{N}}$ for each agent and the 
observation function $O: \mathcal{S} \times \mathcal{A} \mapsto \Delta(\Omega)$ that denotes the conditional probability over the joint observation space $\Omega = \Omega^{1} \times \cdots \times \Omega^{n}$ given state $s' \in \mathcal{S}$ and joint action in the previous timestep $a \in {\mathcal{A}}$.

\paragraph{Coordination.} Accomplishing a shared goal in cooperative settings requires the agents to reach a consensus about how the joint action would improve the performance of all agents. However, since the behavior of one agent influences the reward obtained by other agents, the exploration or sub-optimal action of one agent could overshadow other agents' actions selected at the same time, leading to incorrect reward and value estimation. The misled value estimation could influence the agents' search spaces and result in sub-optimal policies.

\paragraph{Credit Assignment.} Given a shared reward signal in the fully-cooperative setting, the agents could estimate the impact of the joint action on the common reward but lack the ability to estimate the impact of a single agent's action. Even in the centralized training scheme in which agents observe the global state and access to the joint action, the credit assignment problem is difficult to alleviate without strong assumptions on the reward structure~\cite{hu2021rethinking,DBLP:conf/iclr/WangRLYZ21}. Further, the partially observable setting and the decentralized training scheme bring in additional challenges to the credit assignment problem.

\paragraph{Scalability.} In multi-agent scenarios, the agents may take the joint action space $\mathcal{A}$, the joint observation space $\Omega$ and the global state space $\mathcal{S}$ into consideration. The dimension of $\mathcal{A}$ and $\Omega$ grows exponentially w.r.t. the number of agents and the dimension of $\mathcal{S}$ also grows w.r.t. the number of agents. Furthermore, every agent in the environment adds extra complexity for improving the policies. Such properties of multi-agent problems complicate finding and analyzing optimal policies, and the required computation effort grows exponentially w.r.t. the number of agents, leading to the scalability problem of MARL methods.

\begin{table*}[!htbp]
    \centering
    \begin{tabular}{llcl}
        \toprule \multirow{2}{*}{\textbf{Algorithm}} & \multirow{2}{*}{\textbf{Training Scheme}} & \multirow{2}{*}{\textbf{Opponent Awareness}} & \multirow{2}{*}{\textbf{Environment Model Usage}} \\
        & & & \\
        \bottomrule
        MAMBPO~\cite{DBLP:conf/iros/WillemsenCC21} & Centralized& $\times$ & Dyna-style  \\
        CPS~\cite{DBLP:conf/atal/BargiacchiVR21} & Centralized & $\times$ & Dyna-style \\
        MATO~\cite{10.1371/journal.pone.0222215} & Centralized & $\checkmark$ & MPC  \\
        \cite{DBLP:journals/ai/BrafmanT00} & Centralized & $\times$ & Direct Method  \\
        R-MAX~\cite{DBLP:conf/ijcai/BrafmanT01} & Centralized & $\times$ & Direct Method  \\
        M$^3$-UCRL~\cite{DBLP:conf/icml/YangLLZZW18} & Centralized & $\times$ & Direct Method \\
        Tesseract~\cite{DBLP:conf/icml/MahajanSMMGKWZA21} & Centralized & $\times$ & Dynamic Programming \\
        \cite{DBLP:journals/corr/abs-2110-14524} & Centralized & $\times$ & Dynamic Programming \\
        \bottomrule
        AORPO~\cite{DBLP:conf/ijcai/0001WSZ21} & Decentralized & $\checkmark$ & Dyna-style \\
        HPP~\cite{DBLP:conf/corl/WangK0LZITF20} & Decentralized & $\checkmark$ & MPC  \\
        MBOM~\cite{DBLP:journals/corr/abs-2108-01843} & Decentralized & $\checkmark$ & MPC  \\
        MACI~\cite{pretoriusLearningCommunicateImagination2020} & Decentralized & $\times$  & Communication \\
        IS~\cite{DBLP:conf/iclr/KimPS21} & Decentralized & $\checkmark$ & Communication \\
        \bottomrule
    \end{tabular}
    \caption{Classification of model-based MARL algorithms. This classification is first organized according to their training schemes, determined by the settings of possible applications. The next criteria \textit{opponent awareness} refers to whether the agents are aware of and utilize the knowledge of other agents in modeling the dynamics transition or making decisions. We report how the algorithms leverage the learned models in the \textit{Dynamics Model Usage} column.}
    \label{tab: algs}
\end{table*}

\section{
Efficiency of Model-based MARL
}
This section presents the theoretical analyses on the efficiency of model-based MARL. We claim that the efficiency benefits both from leveraging the knowledge of the learned environment model in planning or policy improvement, and alleviating the challenges as mentioned in \Cref{subsec: challenge}.

Model-based methods in single-agent scenarios are known to be efficient, i.e., yield lower sample complexity, both practically~\cite{DBLP:journals/corr/abs-1907-02057} and theoretically~\cite{DBLP:conf/colt/SunJKA019}. We note that the enhanced data efficiency of these model-based methods comes from both leveraging the knowledge of learned environment models and other benefits from learning the models such as the improved stability and the encouraged exploration \cite{DBLP:journals/corr/abs-2006-16712}. Similarly, model-based methods in multi-agent scenarios are proved to have lower sample complexity theoretically, due to the benefits from leveraging the knowledge of the learned environment model. \citet{DBLP:conf/nips/ZhangKBY20} investigated the sample complexity in the two-player discounted zero-sum Markov games. The analyzed model-based MARL method decouples the learning and planning phases: first estimate an empirical model using the collected data and then find the optimal policies in the learned model. Given a generative model, the model-based method that only updates the policy based on the generated data achieves a sample complexity for finding the $\epsilon$-NE policies as $\mathcal{O}(|\mathcal{S}||\mathcal{A}^{1}||\mathcal{A}^{2}|(1-\gamma)^{-3}\epsilon^{-2})$ in the reward-agnostic setting and $\mathcal{O}(|\mathcal{S}|(|\mathcal{A}^{1}|+|\mathcal{A}^{2}|)(1-\gamma)^{-3}\epsilon^{-2})$ in the reward-aware setting, respectively, where $\epsilon$ measures the difference from the accurate Nash Equilibriums. The sample complexities in both cases indicate that the model-based method in multi-agent scenarios is sample-efficient, compared to the sample complexities derived in model-free methods~\cite{DBLP:conf/icml/BaiJ20}.

\citet{DBLP:journals/corr/abs-2110-02355} analyzed the model-based MARL methods in general-sum games from the perspective that the learned dynamics model and learned reward function lead to an approximate game. The Markov perfect equilibrium (MPE) of an approximated game is always an approximate MPE of the original game, denoted as $\alpha$-approximate MPE, where $\alpha$ describes the difference between the $Q$ functions obtained in the approximated game and that in the original game. The model-based MARL methods achieve a sample complexity for finding an $\alpha$-approximate MPE as $\mathcal{O}(|\mathcal{S}||\mathcal{A}|(1-\gamma)^{-2}\alpha^{-2})$, which also shows the superior sample efficiency of model-based MARL methods. Additionally, the robustness analysis establishes an explicit bound for $\alpha$ on the approximation errors of the dynamics model and the reward function. 

Although these theoretical analyses prove the advantages of model-based MARL methods in sample efficiency under limited settings, the study under more general settings is still not investigated. Moreover, the other benefits of model-based MARL methods are not identified.

\section{Algorithms}
\label{sec:alg}

This section provides a review of recent progress in model-based MARL. We first categorize the methods according to their training schemes, for example, centralized training or decentralized training. The classification is illustrated in \Cref{tab: algs}. Then we review each method in detail and analyze them on two aspects:

i) \textit{How the learned models are used?} We focus on the scheme of model usage and the consideration about the properties of multi-agent scenarios, e.g., the interactions among agents when utilizing the learned models. 

ii) \textit{Whether and how the model-based techniques alleviate or sharpen the inherent challenges in MARL?} Although the foremost motivation of model-based methods is to improve the sample efficiency by leveraging the knowledge of learned environment models in planning or policy improvement, a new possibility for resolving the challenges mentioned in \Cref{subsec: challenge} may arise with the learned models. The alleviation of the inherent challenges in multi-agent scenarios could also improve the sample efficiency.

\subsection{Centralized Training}
The direct extension of model-based RL methods in single-agent scenarios to multi-agent scenarios adopts the centralized training scheme. Most of the methods do not consider the properties of multi-agent scenarios but only consider improving the sample efficiency by leveraging the knowledge of the learned environment model in planning or policy improvement, as in single-agent scenarios. 

\paragraph{Dyna-style.} Many recent works have focused on the Dyna-style model-based methods in single-agent scenarios, where the data collected in the real environment is used both to learn environment model and to improve policies. Dyna-style methods in single-agent scenarios show their sample efficiency both practically and theoretically~\cite{DBLP:conf/nips/JannerFZL19}. In terms of which part of the generated data to use when improving the policy, the principles include trusting data from learned models that are estimated to be accurate enough and trusting data that is itself close to the real data. 
MAMBPO~\cite{DBLP:conf/iros/WillemsenCC21} investigates model-based methods in Centralized Training Decentralized Execution (CTDE) paradigm where agents only observe local observations when making decisions, and global information is accessible when agents improving their policies. A centralized environment model $P(s', o', r|s, a)$ is learned to predict the next state, the next joint observation and the current reward, given the current state and current joint action. MAMBPO follows the principle that only the generated data which is close to the real data can be leveraged for policy improvement. Similar to MBPO~\cite{DBLP:conf/nips/JannerFZL19}, the model rollouts begin from the experienced states, and the data generated from the model within $k$ steps is used to improve the policies. MAMBPO can indeed improve the sample efficiency empirically, and alleviate the partial observability and non-stationarity problems by adopting the CTDE paradigm. 

Additionally, with learned dynamics and reward models, extensions of other techniques to multi-agent scenarios could be feasible and efficient. In CPS~\cite{DBLP:conf/atal/BargiacchiVR21}, the agent learns a dynamics model and a reward model to determine which state-action pairs have the higher priority to be updated, and the generated data is also used for policy improvement.

\paragraph{Model Predictive Control.} One typical usage of learned models is planning in the predicted states to select actions. Model Predictive Control (MPC) is a cardinal example, which selects the action with the highest reward in the planning rollouts. MATO~\cite{DBLP:conf/corl/KrupnikMT19} considers the two-agent scenario and aims at reducing the accumulating errors through predicting trajectory segments rather than a single next state, instead of deciding which part of the generated data to use as in the Dyna-style methods. As the fundamental problem of modeling the dynamics in multi-agent scenarios, the interaction between the agents are captured by disentangling the joint model: $\hat{P}(s^{i}_{t+1}, a^{i}_{t}, s^{-i}_{t+1}, a^{-i}_{t}|s^{i}_{t}, s^{-i}_{t}) = \hat{P}^{i}(s^{i}_{t+1}, a^{i}_{t}|s^{i}_{t})\cdot \hat{P}^{-i}(s^{-i}_{t+1}, a^{-i}_{t}|s^{-i}_{t}, s^{i}_{t+1}, a^{i}_{t})$, which means that agent $i$ infers the incoming action and the next state of agent $-i$ based on agent $-i$'s current state, its own current action, and its own predicted next state. After approximating $\hat{P}^{i}$ and $\hat{P}^{-i}$ by variational autoencoders, MATO selects among the latent variables which are the candidates using in the MPC procedure, and determines the best segments. MATO indicates the effectiveness of modeling the interaction among agents in two-agent scenarios. The use of latent variables instead of state-action pairs as candidates in the MPC procedure makes similar methods more feasible to apply in scenarios with more agents, alleviating the scalability challenge.

\paragraph{Direct Method.} The direct method means that the dynamics model is learned based on the data collected from the multi-agent environment and the policy is improved based only on the data generated by the learned models. To the best of our knowledge, \citet{DBLP:journals/ai/BrafmanT00} and R-MAX~\cite{DBLP:conf/ijcai/BrafmanT01} are the earliest model-based algorithms using the direct method in multi-agent scenarios, solving single-controller stochastic games and zero-sum stochastic games respectively. Both methods focus on building a dynamics model and a reward model to resolve the exploration vs. exploitation dilemma. The dynamics model is initialized towards encouraging exploration and transition probability of a state-action pair $(s, a)$ is updated only when $(s, a)$ is encountered enough times, thus balancing exploration and exploitation. M$^3$-UCRL~\cite{DBLP:journals/corr/abs-2107-04050} incorporates model-based techniques in the multi-agent mean-field reinforcement learning method~\cite{DBLP:conf/icml/YangLLZZW18}. An agent using M$^3$-UCRL updates its policy using only the data generated from the dynamics model and the mean-field trajectory model. \citet{DBLP:conf/icml/YangLLZZW18} proved a cumulative regret bound that measures the discrepancy between the cumulative expected reward of the optimal policy and that of the policy generated from the simulated data. These direct model-based methods verify that the learned model could help resolve the exploration vs. exploitation dilemma. In contrast, the direct usage of the learned model may neglect the negative influence of the inaccurate models. 

\paragraph{Dynamic Programming.} Among recent progress in MARL, very few algorithms in the literature use Dynamic Programming (DP), in the sense that the policy and value function are improved considering the state transition probabilities, e.g., the value iteration algorithm updates the policy as $\pi(s) = \operatorname{argmax}_{a} \sum_{s', r}p(s',r|s,a)[r+\gamma V(s')]$~\cite{DBLP:books/lib/SuttonB98}. DP may not be practical for large-scale problems due to its high computation complexity, especially in multi-agent scenarios where the variable spaces grow exponentially with the number of agents. To resolve the scalability challenge, Tesseract~\cite{DBLP:conf/icml/MahajanSMMGKWZA21} shows that the exponential blowup problem in learning policies and value functions can be addressed by accurately representing states and actions using low-rank tensors. This indicates that appropriate classes of models could be found to balance the learnability and expressiveness of policies and value functions. Model-based Tesseract builds an empirical environment model as the sum of low-rank tensors since states and actions are decomposed into low-rank tensors, then evaluates the $Q$ function using DP with this approximate environment model. Based on Tesseract, \citet{DBLP:journals/corr/abs-2110-14524} investigated the generalization of the low-rank dynamics and reward model and shows that low-rank models lead to higher sample efficiency in model-based MARL methods. These methods use dimensionality reduction techniques to describe high dimensional observations from multi-agent tasks as low dimensional feature vectors, reducing the learning complexity in the centralized training scheme. Furthermore, the lower learning complexity and higher sample efficiency provide new possibilities for efficient MARL algorithms, including algorithms in both the centralized training scheme and the decentralized training scheme.

\paragraph{Summary.} The non-stationarity, partial observability, and coordination challenges may not significantly influence the learning problem when methods utilize a centralized training scheme. In contrast, the scalability of the methods becomes increasingly important since the joint action space and the joint observation space are widely used in model-based methods. To resolve the scalability problem, MATO and Tesseract methods perform the planning and improvement in the latent variable spaces or the original variable spaces with reduced dimensions, making it feasible to apply centralized model-based methods into scenarios with multiple agents.
\subsection{Decentralized Training}
Another approach of model-based multi-agent reinforcement learning is to adopt the decentralized training scheme, where the partial observability problem prevails. Additionally, the non-stationarity and coordination problems become more difficult without access to global information. A natural way to deal with these exacerbated problems is through information exchange between the agents, such as the communication methods detailed as below.

\paragraph{Dyna-style.}
Compared to centralized Dyna-style methods, an agent would need to predict other agents' actions to generate model rollouts in decentralized multi-agent training. The actions could be predicted by the learned opponent models or requested from the correspondent agents through communication. AORPO~\cite{DBLP:conf/ijcai/0001WSZ21} investigates improving the sample efficiency in stochastic games, under the setting that agents improve their policies individually but with the ability to communicate with each other. In AORPO, an agent is trained using short rollouts using the decentralized model-based MARL method. From the perspective of a single agent, a return discrepancy upper bound between the expected cumulative reward obtained using AORPO and that obtained with the model-free method is derived, proving the effectiveness of such model-based MARL methods. The theoretical result also indicates the necessity of reducing the opponent models' generalization error. Since the communication cost is one kind of opponent sample complexity that may increase when reducing the opponent models' generalization error, AORPO balances between reducing the generalization error and reducing the opponent sample complexity through the adaptive opponent-wise rollout scheme, each opponent model is used to generate model rollouts for certain number of steps, where the number of steps is determined based on the opponent model's validation error. Actions are requested from the corresponding opponent through communication for the steps afterward. Agents in AORPO learn joint $Q$ functions and the opponent models, alleviating the non-stationarity and coordination problems. Besides, improving policies using the data generated from environment models that predicted joint actions of the agents helps resolve the non-stationarities. However, the opponent models and the joint $Q$ functions require full observability, which rarely holds for real-world applications.

\paragraph{Model Predictive Control.}
With decentralized training scheme, MPC requires a learned environment model and possibly opponent models for an agent to perform decision-time planning. In HPP~\cite{DBLP:conf/corl/WangK0LZITF20},
two agents are presented with a rendezvous task, where they must align their goals without explicit communication. Each agent learns goal-conditioned prediction models to predict the future observations of the agent itself and the other agent given only previous history of its own observation $o_{i}^{t-h:t}$ and a goal $g$, essentially combining the dynamics model and the opponent models. The models are pretrained with supervised learning on datasets collected with existing controllers. The higher-level HPP planner then uses the cross-entropy method \cite{CEM} to select the best goal for every $T_h$ timesteps during execution after receiving the history of observations of all agents for these timesteps.
During execution, the high-level HPP planner updates its beliefs over potential rendezvous points and adjusts the goal for the low-level controller accordingly, which leads to agents reaching a common rendezvous. 
The fact that HPP plans over the goal space instead of the action space allows more effective planning due to the lowered dimensionality of the goal space. It also helps accelerate training because goals can be instructive throughout multiple timesteps.

To account for unseen or adapting agents, MBOM~\cite{DBLP:journals/corr/abs-2108-01843} investigates the learning process of opponents by modeling it as recursively fine-tuning pretrained opponent models. Each stage of fine-tuning is done by calculating the best response of the fine-tuned model of the previous stage. Specifically, before each interaction, a MBOM agent generates $M$ fine-tuned models $\{\phi\}_{i=0}^{M-1}$ and executes a mixed policy $\pi_{\text{mix}}(\cdot|s) = \sum_{i=0}^{M-1} \alpha_i\pi(\cdot|s; \phi_i)$ with Bayesian mixing. This decision-time planning, with opponent models of varying reasoning levels, allows MBOM to relax previous restrictions on opponent modeling and potentially compete or coordinate with changing or unseen opponents. Decision-time planning by recursive reasoning also increases sample efficiency by generating high-quality responses. However, MBOM has no guarantee about the convergence of the recursive reasoning of opponents' actions and requires full observability.


\paragraph{Communication Methods.}
Communication methods benefit from environment models since the messages are thus able to convey agents' predictions about the long-term future rather than the immediate partial observations, which facilitates agents to reach a common consensus. 
Communication protocols can reduce dynamics sample complexity since the agent can leverage the knowledge of the messages from other agents, at the cost of increasing opponent complexity, i.e., the communication cost.
MACI~\cite{pretoriusLearningCommunicateImagination2020} leverages a communication protocol that allows agents to communicate with their neighbors for multiple rounds before deciding on a joint action. The message for each round encodes each agent's imagined rollouts generated using a recurrent dynamics model which takes as input agent's local observation and the incoming message from the neighbors. The proposed communication protocol is differentiable and thus the encoder, and the dynamics model can be learned with end-to-end training. The dynamics model enables communication for decision-time planning, which acts to alleviate the non-stationarity and partial observability problem since the communication allows agents to share their observations and policies implicitly. In terms of sample complexity, while multiple communication rounds increase opponent sample complexity to some degree, dynamics sample complexity is reduced because the multiple rounds of exchanging information accelerate the consensus-reaching process. 

Similarly, Intention Sharing (IS)~\cite{DBLP:conf/iclr/KimPS21} proposes a communication protocol that also encodes into the message an agent's imagined trajectory of multiple steps. From the perspective of agent $i$, the imagined trajectory is generated by performing rollouts with the opponent model that predicts other agents' actions $a^{-i}$ based on its local observation $o^{i}$, and a dynamics model $P({o^{i}}'| o^{i}, a^{i}, \hat{a}^{-i})$ which takes the predicted $\hat{a}^{-i}$ as input. IS further incorporates an attention module to focus on encoding the more important parts of the imagined trajectory. Different from MACI, IS utilizes both a dynamics model and opponent models to generate imagined rollouts. With opponent models, the dynamics model takes as input the joint action instead of the action of a single agent, further addressing the non-stationarity and coordination problems since interactions among agents are now considered. The sample complexity is greatly affected by the usefulness of the encoded message, which is why IS adopts the attention module to focus on the more critical parts of the trajectory.
\paragraph{Summary.}
In the decentralized training scheme, the non-stationarity, partial observability, and coordination problems become more critical. In contrast, scalability is less concerned than that in the centralized training scheme. Opponent modeling is often introduced in decentralized model-based methods to alleviate the non-stationarity and coordination problems to explicitly model the decision-making process of other agents. In terms of model-based methods, opponent modeling makes Dyna-style and MPC methods possible in decentralized settings, which require predicting the state transition and other agents' actions when performing model rollouts. However, modeling opponents in partially observable environments remains an open problem~\cite{DBLP:journals/ai/AlbrechtS18}.
Additionally, under communication protocols that utilize the environment model, agents can share their prediction of the long-term future instead of only the intention at the current step, which assists agents in effectively reaching a consensus. Effective communication protocols also help with increasing dynamics sample efficiency because a consensus is quicker to form under the effective exchange of long-term information.
\section{Conclusions and Future Directions}

This paper takes a review of the recent progress of theoretical analyses, algorithms, and applications in model-based MARL. The classification of the algorithms is organized on three dimensions, i.e., the training scheme, the opponent awareness, and the environment usage. We discuss how the learned environment models, possibly together with opponent models, are leveraged in detail and the influence of these model-based methods on the inherent challenges of multi-agent scenarios. Model-based MARL algorithms are still under-researched: some fundamental problems of model-based methods, including the model inaccuracy, the mismatch between the environment model, the value function, and so on, as well as multi-agent properties, for instance, the interaction among agents and the credit assignment problem in fully cooperative settings, have not been extensively investigated. We summarize the pros and cons of the algorithms in both centralized training and decentralized training schemes. It is worth generalizing the techniques and principles in both model-based RL and model-free MARL for the advantages in sample efficiency and the possibility to resolve the inherent challenges in multi-agent scenarios. More specifically, since model-based MARL is a new and promising branch of MARL, there are many future directions in both the centralized training scheme and the decentralized training scheme. We now provide several research directions that have the potential to push the future development of this field based on our knowledge.
\paragraph{Scalability in Centralized Model-based MARL.}
The scalability becomes a major challenge in centralized training model-based algorithms, which requires large amounts of computational resources and increases the learning complexity in both learning and leveraging environment models. The scalability problem limits the incorporation of effective methods such as Dynamic Programming and Monte Carlo Tree Search~\cite{browne2012survey} into model-based MARL. Reducing the joint variables' dimensions shows its promise for effective model learning and usage~\cite{DBLP:conf/icml/MahajanSMMGKWZA21,DBLP:conf/corl/KrupnikMT19}.

\paragraph{Decentralized Model-based MARL with Opponent Modeling.} 
Scenarios where agents improve their policies individually, are often characterized by partial observability, which makes resolving non-stationarity and coordination problems more difficult. Although suffering from the partial observability problem, opponent modeling is a natural approach for alleviating these challenges and is necessary for capturing the interaction among agents when modeling the environment. In addition, the environment model's learnability increases with opponent modeling since the learning objective could be decoupled into learning a dynamics model with stabilized opponents and learning accurate opponent models.

\paragraph{Communication with Learned Models.}
Messages encoded with future trajectories predicted from environment models outperform those encoded with only current-step information regarding capturing agents' intentions and leading agents to common consensuses. Extensions of techniques developed in communication-based MARL into model-based MARL are worth investigating, such as automatic learning of communication protocols and influence measurement among agents.

We expect that this paper presents the advantages and potential of model-based MARL and reveals its opportunities and limitations. In the foreseeable future, we hope that the community could be inspired by this paper and encouraged for further developments in this attractive, necessary, and young field of research.

\bibliographystyle{named}
\bibliography{ref}

\end{document}